\documentclass[prl,notitlepage,twocolumn,tightenlines,nobibnotes,superscriptaddress]{revtex4-1}

\usepackage{amsmath,textcomp,graphicx,color}
\newcommand{\bra}[1]{\left\langle #1 \right\vert}
\newcommand{\ket}[1]{\left\vert #1 \right\rangle}
\newcommand{\pcite}[1]{\kern -0.1em~~\cite{#1}}

\begin{document}
\title{Enhancing quantum transport in a photonic network using controllable decoherence}

\author{Devon N. Biggerstaff}
\affiliation{Centre for Engineered Quantum Systems and Centre for Quantum Computation and Communication Technology, School of Mathematics and Physics, The University of Queensland, Brisbane QLD 4072, Australia}
\author{Ren\'e Heilmann}
\affiliation{Institute of Applied Physics, Abbe Center of Photonics, Friedrich-Schiller Universit\"at Jena, Max-Wien-Platz 1, D-07743 Jena, Germany}
\author{Aidan A. Zecevik}
\affiliation{Centre for Engineered Quantum Systems and Centre for Quantum Computation and Communication Technology, School of Mathematics and Physics, The University of Queensland, Brisbane QLD 4072, Australia}
\author{Markus~Gr\"afe}
\affiliation{Institute of Applied Physics, Abbe Center of Photonics, Friedrich-Schiller Universit\"at Jena, Max-Wien-Platz 1, D-07743 Jena, Germany}
\author{Matthew A. Broome}
\altaffiliation{Current address: Centre for Quantum Computation and Communication Technology, School of Physics, University of New South Wales, Sydney NSW 2052, Australia}
\affiliation{Centre for Engineered Quantum Systems and Centre for Quantum Computation and Communication Technology, School of Mathematics and Physics, The University of Queensland, Brisbane QLD 4072, Australia}
\author{Alessandro Fedrizzi}
\affiliation{Centre for Engineered Quantum Systems and Centre for Quantum Computation and Communication Technology, School of Mathematics and Physics, The University of Queensland, Brisbane QLD 4072, Australia}
\author{Stefan~Nolte}
\affiliation{Institute of Applied Physics, Abbe Center of Photonics, Friedrich-Schiller Universit\"at Jena, Max-Wien-Platz 1, D-07743 Jena, Germany}
\author{Alexander Szameit}
\affiliation{Institute of Applied Physics, Abbe Center of Photonics, Friedrich-Schiller Universit\"at Jena, Max-Wien-Platz 1, D-07743 Jena, Germany}
\author{Andrew G. White}
\affiliation{Centre for Engineered Quantum Systems and Centre for Quantum Computation and Communication Technology, School of Mathematics and Physics, The University of Queensland, Brisbane QLD 4072, Australia}
\author{Ivan Kassal}
\email{Email: i.kassal@uq.edu.au}
\affiliation{Centre for Engineered Quantum Systems and Centre for Quantum Computation and Communication Technology, School of Mathematics and Physics, The University of Queensland, Brisbane QLD 4072, Australia}

\begin{abstract}
Transport phenomena on a quantum scale appear in a variety of systems, ranging from photosynthetic complexes to engineered quantum devices. It has been predicted that the efficiency of quantum transport can be enhanced through dynamic interaction between the system and a noisy environment. We report the first experimental demonstration of such environment-assisted quantum transport, using an engineered network of laser-written waveguides, with relative energies and inter-waveguide couplings tailored to yield the desired Hamiltonian. Controllable decoherence is simulated via broadening the bandwidth of the input illumination, yielding a significant increase in transport efficiency relative to the narrowband case.
We show integrated optics to be suitable for simulating specific target Hamiltonians as well as open quantum systems with controllable loss and decoherence.
\end{abstract}

\maketitle

Recent research into photosynthetic antenna complexes  has shown evidence of coherence in excitonic energy transport~\cite{engel07,engel_2010,scholes_2010,Wong:2012jd}, 
despite the noisy cellular environment in which such complexes are found. Indeed, environmental decoherence has been credited with increasing the efficiency of transport through these systems, an effect known as environment-assisted quantum transport (ENAQT) ~\cite{rebentrost2009eaq} or dephasing-assisted transport~\cite{plenio2008,chin2010}. While ENAQT has been the subject of many theoretical studies---whether in the photosynthetic context~\cite{Mohseni:2008gp,Cao:2009vs,Caruso:2009ib,Wu:2010bg,whaley_2010,alexandra_2010,Wu:2013ig,Cleary:2013ia,leon2014} or in other nanoscale transport systems~\cite{semiao2010,Scholak:2011dy,Scholak:2011fv,lim2014}---and despite its potential for improving transport in artificial quantum systems, it has so far never been directly observed.

We use an integrated photonic simulator to demonstrate the first implementation of ENAQT. Our simulator was fabricated using femtosecond-laser direct writing, which allows waveguides to be drawn directly into glass using a focused pulsed laser. This permits the creation of three-dimensional waveguide arrays, as well as precision and repeatability in engineering interactions~\cite{szameit2007cwa,szameit2010jpb,marshall2009lww,OwensQW,Corrielli2013,poulios,crespi2015fano}. We used control over the wavelength and bandwidth of the guided light to 
simulate effective decoherence, thereby enhancing transport efficiency.

\begin{figure*}[hbt!]
\centering
\includegraphics[width=\textwidth]{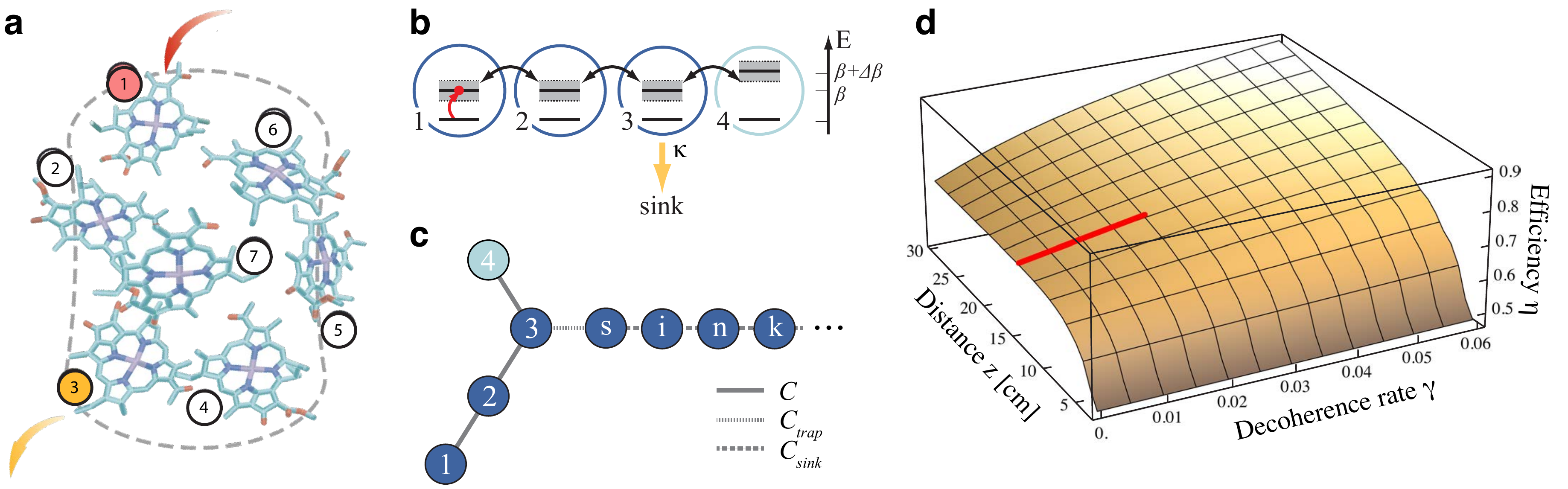}
\caption{Environment-assisted quantum transport. 
\textbf{a} Photosynthetic antenna complexes are networks of chlorophylls that collect and transfer solar energy. A well-studied example is the Fenna-Matthews-Olson complex of green sulphur bacteria, here depicted as a network of 7 sites that transports excitation energy from initial site 1 to target site 3 (adapted with permission from ref. \onlinecite{lambert2013quantum}). Simulations have suggested that this transport may be enhanced by decoherence~\cite{rebentrost2009eaq,plenio2008,chin2010}.
\textbf{b} We simulate an instance of ENAQT on a lattice of 4 sites, with site 1 initially excited and site 3 the target. If the detuning $\Delta\beta$ of site 4 equals $C$, one of the system eigenmodes has no occupancy at site 3 and cannot couple to the sink; by broadening the levels, decoherence breaks the condition $\Delta\beta=C$, allowing all eigenmodes to couple to the sink and thus increasing transport efficiency. 
\textbf{c} Our simulator consists of 4 coupled waveguides arranged as shown (cross-section). The sink is modelled with a large array of closely coupled waveguides that transport light away from the main 4 waveguides. At the central wavelength $\lambda_0$, waveguide 4 has propagation constant $\beta+\Delta\beta$, while the others have propagation constant $\beta$.
\textbf{d} Theoretical expectation of transport enhancement for this system, as a function of simulation length $z$ and decoherence strength $\gamma$. The red bar indicates the region explored experimentally.
\label{fig:enaqt}}
\end{figure*}

We consider a single excitation on a network of $N$ coupled sites, governed by a tight-binding Hamiltonian~\cite{rebentrost2009eaq}
\begin{equation}
H = \sum_{m=1}^{N} \varepsilon_m \ket{m}\!\bra{m} + \sum_{n<m}^{N} V_{mn} \left( \ket{m}\!\bra{n}+\ket{n}\!\bra{m}  \right),
\label{eq:tight-binding}
\end{equation}
where $\ket{m}$ denotes the excitation being localised at site $m$, $\varepsilon_m$ the energy of that site, and $V_{mn}$ the coupling between sites $m$ and $n$. Although ENAQT can occur on an ordered lattice where all the energies $\varepsilon_m$ are equal~\cite{kassal2012njp}, transport enhancement was first explained in disordered systems, which we consider here.

We are interested in the efficiency of transport from a particular initial site to a particular target site, where the excitation is trapped. In the case of a photosynthetic complex (figure~\ref{fig:enaqt}a), trapping describes the transfer of excitons to a reaction center, where they drive charge separation. It can be modelled as irreversible coupling of the target site to a sink at rate $\kappa$ (figure~\ref{fig:enaqt}b). The efficiency is then the probability of finding the exciton in the sink after some particular time. 

ENAQT occurs when decoherence increases the trapping probability over the fully coherent case. Decoherence results from coupling of a quantum system to inaccessible degrees of freedom. For example, in photosynthetic antenna complexes, the energies of chromophores are coupled to molecular vibrations; tracing out this environment results in decoherence in the excitonic subspace.

In the absence of decoherence, energetic disorder tends to localise the wavepacket through processes such as destructive interference or Anderson localisation~\cite{anderson1958,lahini_anderson_2008,segev2013}, thus preventing it from reaching the target. Since these are coherent processes, they are diminished by decoherence, possibly resulting in enhancement of transport efficiency to the target site. An alternative but ultimately equivalent view of ENAQT is that eigenstates of $H$ are stationary, making it difficult to reach the target if the initial and target sites differ in energy. Incoherent processes, however, permit transitions between the eigenstates of $H$, yielding greater mobility. 

The first theoretical explorations of ENAQT focused on the case where the decoherence takes the form of site-independent, Markovian, pure dephasing~\cite{plenio2008,rebentrost2009eaq}. Although the decoherence in our simulation is neither site-based nor Markovian, this is not an obstacle to ENAQT;~\cite{mohseni2014,chen2011} transport efficiency can be enhanced as long as the decoherence allows population transfer between otherwise-stationary eigenstates.

We simulate ENAQT in an array of coupled single-mode optical waveguides obeying the equation~~\cite{perets_realization_2008,longhi2009review}
\begin{equation}
i\frac{\partial}{\partial z}a_{m}^{\dagger}(z)=\beta_{m}a_{m}^{\dagger}(z)+\sum_{n\ne m}^{N}C_{mn}a_{n}^{\dagger}(z),
\label{eq:waveguides}
\end{equation}
where the light is propagating in the $z$-direction, $a_m^\dagger(z)$ is a creation operator for a photon in waveguide $m$ at position $z$, and $\beta_m$ and $C_{mn}$ are respectively the propagation constants of the waveguides and the couplings between them. The former are determined by the waveguides' refractive index profiles, while the latter also depend on the separations between them. Light propagation governed by this Schr\"odinger-like equation directly simulates evolution under $H$, with $C_{mn}$ replacing $V_{mn}$ and the $\beta_m$ replacing $\varepsilon_m$.
We can thus simulate different Hamiltonians by controlling the number, position, and refractive indices of the waveguides.

The intrinsic stability of laser-written waveguide arrays renders decoherence challenging to simulate.
One approach is to stochastically modulate the index of refraction along each waveguide~\cite{levi2012natphys}. Though every realization with a particular longitudinal index profile will be fully coherent, decoherence can be simulated by averaging over the recorded optical outputs from many realizations in post-processing. This approach was recently used to simulate decoherence-enhanced navigation of a maze~\cite{caruso2015maze}.

By contrast, we simulate decoherence by averaging over the results from a single array illuminated with many optical wavelengths. Although each individual wavelength propagates through the waveguide array coherently, decoherence can be achieved using broadband illumination and a single output intensity measurement which does not resolve wavelength. In other words, the wavelength degree of freedom is traced out yielding a partially mixed state. Similar approaches are well established in optics, where for instance thick birefringent quartz plates followed by wavelength-insensitive measurements have been used to decohere the polarization states of single photons~\cite{kwiat2000dfs,xu2010decoherence}.

\begin{figure*}[hbt]
\centering
\includegraphics[width=\textwidth]{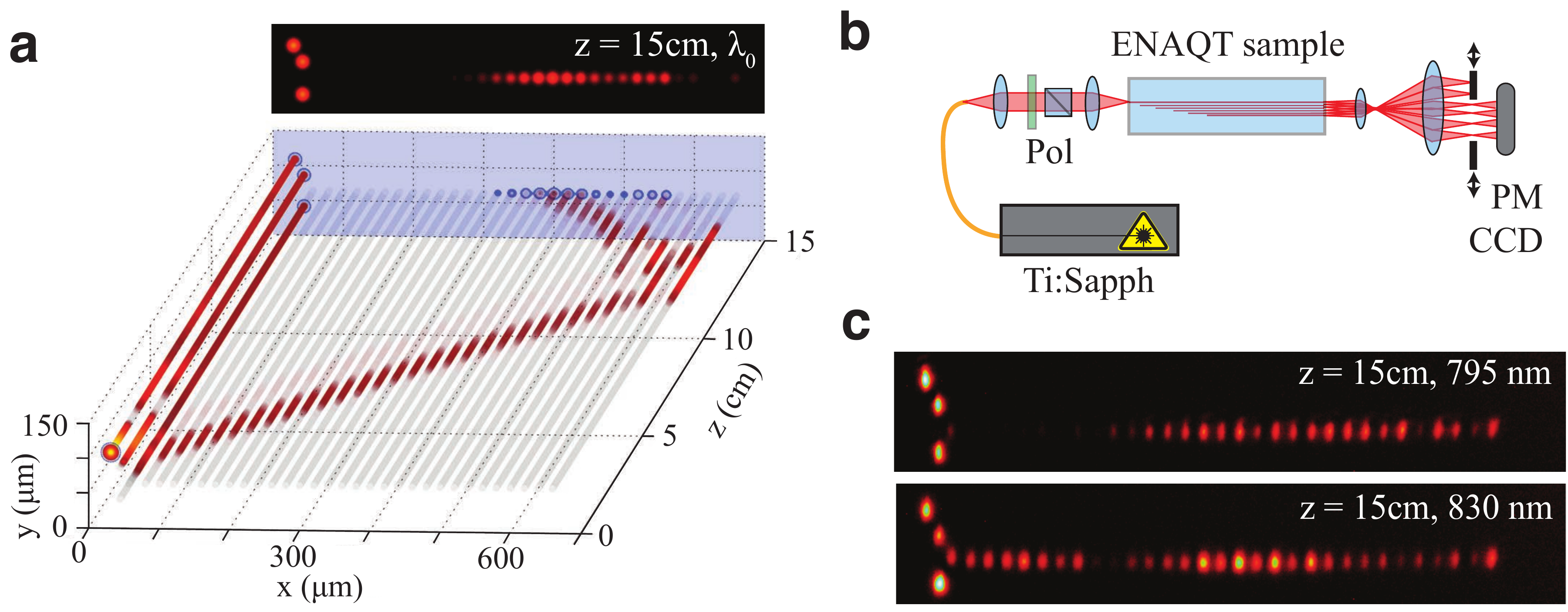}
\caption{Experimental setup and waveguide design.
\textbf{a} Predicted dynamics of light in our waveguide array, as a function of $z$. The inset shows the predicted device output distribution for input at wavelength $\lambda_0$. The sink array is sufficiently long that light reflecting from the far boundary fails to couple back into the system waveguides during the simulation.
\textbf{b} A fibre-coupled, tuneable Ti:sapphire laser in quasi-cw mode undergoes polarisation control (Pol) before it is imaged into the sample using a 15~mm focal-length aspheric lens. The output is imaged via a 14${\times}$ telescope onto a variable slit, which collectively measures the total intensity output from the system, bath, or all the waveguides using a large-area power-meter (PM). Alternatively, the output can be imaged onto a CCD camera for alignment and diagnostics.
\textbf{c} CCD images of the output after $15$~cm when illuminated at $\lambda_0=792.5$~nm and at $830$~nm. As designed, the light in the system waveguides is evenly distributed at $\lambda_0$ apart from the target site, which is dark. In contrast, the target site is much brighter at $830$~nm, indicating that more light will couple into the sink---a sign of ENAQT.
\label{fig:setup}}
\end{figure*}

As an example of our approach to simulating decoherence, we consider two uncoupled waveguides $a$ and $b$ that have a difference $\Delta \beta$ in their propagation constants at a wavelength $\lambda_0$. As light propagates along the waveguides, it will accumulate a phase difference $z\Delta \beta$ between them. Broadband illumination can then be seen to cause effective decoherence on a length scale comparable to the illumination coherence length: Any initial coherence $\rho_{ab}$ between the waveguides will decay:
\begin{equation}
\rho_{ab}(z)=\rho_{ab}(0) \; g^{(1)} \! \left(\frac{z\Delta\beta\, \lambda_0}{2\pi c}\right),
\label{eqn:g1}
\end{equation}
where $g^{(1)}(t)$ is the normalised first-order temporal correlation function of the light, which---being proportional to the Fourier transform of the spectrum---decays to zero faster for spectrally broader illumination. The strength $\gamma$ of the decoherence can be quantified as the inverse of the optical coherence length, $\gamma = \left(\frac{2\pi c}{\Delta\beta \,\lambda_0} \int_{-\infty}^{\infty} |g^{(1)}(\tau)|^2 \,d\tau \right)^{-1}$, and is usually proportional to the FWHM bandwidth $\Delta\lambda$. For a uniform distribution centered at $\lambda_0$,
\begin{equation}
\gamma = \frac{\Delta \beta\, \Delta\lambda}{ 2\pi\lambda_0 }.
\label{eqn:gamma}
\end{equation}

In the case of coupled waveguides, wavelength dependence affects couplings in addition to the propagation constants. The resulting decoherence will therefore not only have characteristics of pure dephasing (as in equation~\eqref{eqn:g1}), but will also include off-diagonal terms. However, this kind of decoherence can also result in ENAQT. Indeed, the off-diagonal terms imply that our decoherence could, in principle, be used to simulate ENAQT in ordered systems~\cite{kassal2012njp}, but the effect would be weaker because of the absence of the pure-dephasing contribution.

For our simulation we chose a network of four waveguide sites---arranged as in figure~\ref{fig:enaqt}c---because it is one of the smallest systems in which ENAQT is possible and because it can give significant enhancements even with relatively weak decoherence. Waveguide 1 was the input, waveguide 3 the target, and the sink consisted of a long linear array of tightly coupled waveguides. The coupling between waveguides in this sink was significantly higher than between the four main waveguides, so that any light entering the sink from waveguide 3 was largely transported away~\cite{longhi2006,biagioni2008,delanty2012novel}. The sink need only be long enough to prevent light reflected from the far end from returning into the main simulation waveguides.

We need to choose propagation constants and separations among the four main waveguides in order to best approximate this Hamiltonian:
\begin{equation}
H = \begin{pmatrix}
	\beta & C & 0 & 0 \\
	C & \beta & C & 0 \\
	0 & C & \beta & C \\
	0 & 0 & C & \beta+\Delta\beta \\
\end{pmatrix},
\label{eqn:H}
\end{equation}
where all parameters depend on the wavelength $\lambda$ of the input light. Due to their wide separations, couplings between non-neighbouring sites are negligible ($<5\%$ of neighbouring-site couplings). Our simulator is designed so that, at our central simulation wavelength $\lambda_0$,
\begin{equation}
\Delta\beta(\lambda_0)=C(\lambda_0).
\label{eqn:condition}
\end{equation}
In this case, one of the eigenstates of $H$ has no support on site 3, $\ket{\psi_1}=(-1,-1,0,1)/\sqrt{3}$, while
the remaining three eigenstates all have substantial support on site 3.
Because $\ket{\psi_1}$ cannot couple to the sink---at least at $\lambda_0$---the maximum trapping efficiency at infinite time is $\eta=1-\langle1|\psi_1\rangle=2/3$.

Considering the wavelength dependence of $H$ provides a different way to see the decohering effects of broadband illumination.
Due to the dependence of $\beta$ and $C$ on the wavelength, equation~\eqref{eqn:condition} only holds at a particular wavelength $\lambda_0$. At other wavelengths, $\ket{\psi_1}$ will have some support on waveguide 3 and thus be able to couple to the sink, increasing the efficiency above $2/3$, as shown in figure~\ref{fig:enaqt}d. Unlike in other examples of ENAQT~\cite{rebentrost2009eaq}, the efficiency increases monotonically with the strength of decoherence, meaning there is no optimal level of decoherence in this model.

Based on measurements of couplings and propagation constants in isolated pairs of waveguides (see Methods), we selected the following design parameters: $\Delta\beta=C=1.0~\mathrm{cm}^{-1}$, $C_\mathrm{trap}= 1.5~\mathrm{cm}^{-1}$, and $C_\mathrm{sink}=1.75~\mathrm{cm}^{-1}$ (see figure~\ref{fig:enaqt}c). Figure~\ref{fig:setup}a shows numerical modelling of light propagation given these parameters.
Although these were designed for a center wavelength of 800~nm, variations in the implementation of the waveguide parameters resulted in equation~\ref{eqn:condition} being satisfied at $\lambda_0=792.5$~nm. 

\begin{figure*}[hbt!]
\centering
\includegraphics[width=\textwidth]{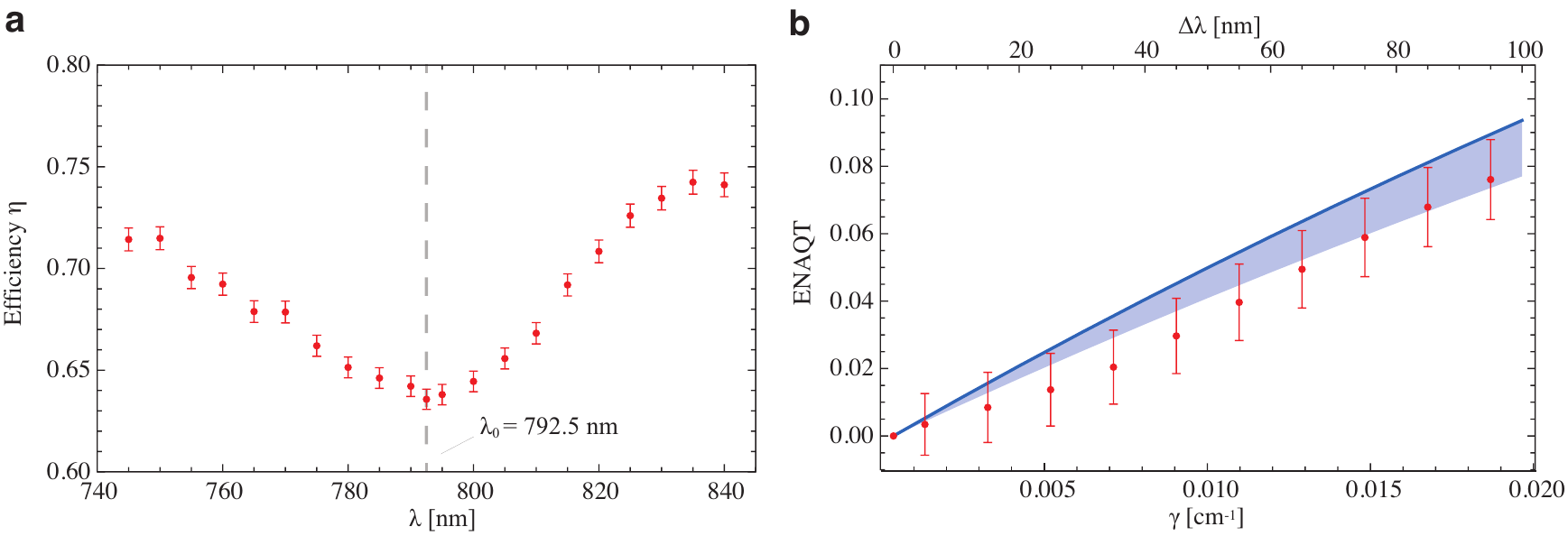}
\caption{Magnitude of observed ENAQT. 
\textbf{a} Transport efficiency---the portion of light that makes it to the sink---as a function of wavelength. The minimum efficiency at $\lambda_0=792.5$~nm is $\eta=0.636\pm 0.002$, slightly less than the theoretical infinite-time limit of $2/3$. The error bars are standard deviations caused by imperfect repeatability in coupling light into the sample and laser power fluctuations. 
\textbf{b} ENAQT---the relative increase in the efficiency over that at $\lambda_0$---as a function of the optical bandwidth (top horizontal axis) and corresponding decoherence strength $\gamma$ (bottom horizontal axis). The red points are obtained by averaging the measured efficiencies over a uniform broadband spectrum with width $\Delta\lambda$ and centered at $\lambda_0$.
The blue line represents the theoretical ENAQT, calculated based on the model in figure~\ref{fig:enaqt}b and containing no free parameters (it is the cross-section of figure~\ref{fig:enaqt}d along the red line segment). The shaded region represents possible ENAQT if $\Delta\beta(\lambda_0)$ deviates from $C(\lambda_0)$ by up to 10\%.
\label{fig:results}}
\end{figure*}

We measured the efficiency using narrowband light (less than 1~nm bandwidth and always horizontally polarised for consistency) from a tuneable Ti:sapphire laser (Spectra-Physics Tsunami) in quasi-cw mode (figure~\ref{fig:setup}b). 
The output was imaged using a custom-built $14{\times}$ magnifying telescope and the optical power was measured using a large-area power-meter after isolating either the system or sink waveguides using a variable slit. Examples of the output distribution are given in figure~\ref{fig:setup}c, showing the significant difference between illumination at $\lambda_0$ and an off-center wavelength.
Figure~\ref{fig:results}a shows the measured efficiency (fraction of light output in the sink modes) for wavelengths ranging from 745 to 835~nm.

ENAQT---shown in figure~\ref{fig:results}b---is the average enhancement in efficiency over the spectral band of interest, relative to the efficiency at $\lambda_0$,
\begin{equation}
\text{ENAQT}=\frac{\left\langle\eta(\lambda)\right\rangle-\eta(\lambda_0)}{\eta(\lambda_0)},
\label{eqn:enaqt}
\end{equation}
where the average $\langle\cdots\rangle$ over $\lambda$ is taken over the top-hat  distribution on $(\lambda_0-\frac12 \Delta\lambda,\lambda_0+\frac12 \Delta\lambda)$. As predicted, it is an increasing function of the bandwidth, i.e., of the decoherence strength $\gamma$. The highest ENAQT observed was $(7.6 \pm1.2)\%$ at a bandwidth of $95$~nm. We thus demonstrated that coherent transport in a coupled, statically disordered system can be enhanced through decoherence.

The theoretical prediction in figure~\ref{fig:results}b contains no free parameters. It is a simulation of the dynamics under the simulated Hamiltonian $H$, together with trapping from site 3 at the rate $\kappa$ discussed in Methods and pure dephasing between waveguide $4$ and the other three waveguides at rate $\gamma$. The disagreement between theory and experiment is small considering the number of possible contributing factors. These include the off-diagonal decoherence when the waveguides are coupled, the fact that the trapping is not perfectly exponential, errors in the measurements of the coupling constants, optical losses, and error in satisfying equation~\ref{eqn:condition}. As an example, the shaded band in figure~\ref{fig:results}b represents the error that would arise if $\Delta\beta(\lambda_0)$ deviated from $C(\lambda_0)$ by up to 10\%.

In our experiment, the magnitude of ENAQT was limited by the maximum achievable decoherence, which at $\gamma = 0.02$~cm$^{-1}$ was small compared to the inverse of the propagation length.  We were limited by two components of equation \eqref{eqn:gamma}:
 the tunability of our laser limited $\Delta\lambda$, while $\Delta\beta$ was limited (via equation \eqref{eqn:condition}) by the need to keep $C$ small enough to stay in the tight-binding approximation.
These are not fundamental limitations, and future improvements that increase $\gamma$ would result in significantly larger transport enhancement; as shown in figure~\ref{fig:enaqt}d, the transport efficiency of this model can get arbitrarily close to 1 for sufficiently long propagation distances and decoherence strengths.
Stronger decoherence would also allow ENAQT to be observed in networks that are less sensitive to decoherence than our model.

Our results demonstrate that integrated photonics are well-suited for simulating open quantum systems, capable of implementing a disordered target Hamiltonian with controllable loss. The technique of using broadband excitation to introduce tuneable levels of decoherence ensures that photonics can simulate open quantum systems; 
previously, low intrinsic noise in photonic devices rendered decoherence difficult to realize in integrated optics, particularly without averaging over results from many different device realizations.
Our approach will not only improve the controllability of photonic quantum simulation, 
but also aid in the experimental optimisation of transport in other engineered quantum systems.

\section*{Methods}

Our experimental setup is shown in figure~\ref{fig:setup}b. The waveguides were fabricated in high-purity fused silica (Corning 7980) using a laser direct-write technique whereby Ti:sapphire laser pulses are tightly focused into the sample, which is then translated in three dimensions to yield continuous regions of positive refractive index change which act as waveguides~\cite{szameit2010jpb}. To obtain single-mode waveguides with the desired propagation and coupling characteristics, laser pulses of duration 150~fs, energy 400~\textmu J, central wavelength 800~nm, and repetition rate 100~kHz were focused 400~\textmu m below the surface using a 40${\times}$ microscope objective and translated at 75~mm/min (for waveguides other than 4).

Permanent index changes result in the focus, yielding elliptical, vertically oriented, single-mode waveguides with modes approximately $17{\times}19$~{\textmu}m in size at 800~nm. Waveguide 3 and the sink waveguides are in a plane parallel to the surface, while angles of 120\textdegree{} between this plane and the other system waveguides minimise non-neighbour coupling. Because couplings between elliptical waveguides are dependent on angular orientation~\cite{szameit2007cwa}, we determined the couplings by writing pairs of waveguides oriented at the specified angles. Couplings as a function of separation were determined by writing the pairs at different separations and measuring the output intensities after a known propagation length when only one is optically excited.

The final waveguide separations were chosen to ensure that the tight-binding approximation is maintained, that $C$ can be matched by $\Delta\beta$, that $C_\mathrm{sink} > C_\mathrm{trap} > C$, and that the number of sink modes remains manageable. The system-sink population transfer can be modelled as a constant effective rate~~\cite{longhi2006} $\kappa=2x^2(1-x^2)^{-1/2}$ if $x\equiv C_\mathrm{trap}/C_\mathrm{sink} \ll 1$. In our case, although $x=0.86$ meant that the decay was not perfectly exponential, it was nevertheless effectively irreversible, considering that sufficiently many sink modes were present to prevent reflection from the far end and back into the main waveguides.

For waveguide 4, the writing translation speed was decreased to increase the propagation constant. In a pair of waveguides with mismatched propagation constants, the maximum power transfer depends only on the ratio $\Delta\beta/C$. We used this fact to determine the translation speed decrease necessary to set $\Delta\beta=C$ at 800~nm. Due to the measurement uncertainty (about $8\%$) and day-to-day variability in the waveguide writing process, several arrays were written with slightly different writing speeds for waveguide 4 to ensure that one set would yield $\Delta\beta=C$ in our desired wavelength range. In the best sample, the writing speed was decreased by 12\%.

\section*{Acknowledgements}

DNB, AAZ, MAB, AF, AGW, and IK were supported by Australian Research Council (ARC) 
Centres of Excellence for Engineered Quantum Systems (CE110001013)
and Quantum Computation and Communication Technology (CE110001027).
RH, MG, SN, and AS thank the German Ministry of Education and Research
(Center for Innovation Competence program, 03Z1HN31), 
the Thuringian Ministry for Education, Science and Culture (Research group Spacetime, 11027-514), 
the Deutsche Forschungsgemeinschaft (NO462/6-1), and the 
German-Israeli Foundation for Scientific Research and Development (1157-127.14/2011).
AF and IK acknowledge ARC Discovery Early Career Researcher Awards (DE130100240 and DE140100433). 

%\bibliography{enaqt.bib}
%\bibliographystyle{apsrev.bst}

\end{document}